\documentclass[pra,twocolumn,showpacs,amsmath,amssymb]{revtex4}
\usepackage{graphicx}
\usepackage{amsfonts}
\usepackage{amsmath}
\usepackage{amssymb}
\usepackage{color}
\usepackage{bm}
\usepackage{bbm}
\usepackage{enumerate}

\graphicspath{{figures/}} 
\usepackage{amsfonts, amsmath, amsthm, amssymb} 
\usepackage{array}
\usepackage[colorlinks,bookmarks=false,citecolor=blue,linkcolor=red,urlcolor=blue]{hyperref}

\newcommand{\be}{\begin{equation}}
\newcommand{\bee}{\begin{equation*}}
\newcommand{\ee}{\end{equation}}
\newcommand{\eee}{\end{equation*}}
\newcommand{\bearre}{\begin{eqnarray*}}
\newcommand{\eearre}{\end{eqnarray*}}
\newcommand{\bearr}{\begin{eqnarray}}
\newcommand{\eearr}{\end{eqnarray}}

\begin{document}

\title{Quantum correlations, separability and quantum coherence length in equilibrium many-body systems}

\author{Daniele Malpetti$^1$
 and Tommaso Roscilde$^{1,2}$
 }

\affiliation{$^1$ Laboratoire de Physique, CNRS UMR 5672, Ecole Normale Sup\'erieure de Lyon, Universit\'e de Lyon, 46 All\'ee d'Italie, 
Lyon, F-69364, France}
\affiliation{$^2$ Institut Universitaire de France, 103 boulevard Saint-Michel, 75005 Paris, France}

\date{\today}

\begin{abstract}
Non-locality is a fundamental trait of quantum many-body systems, both at the level of pure states, as well as at the level of mixed states. Due to non-locality, mixed states of any two subsystems are correlated in a stronger way than what can be accounted for by considering correlated probabilities of occupying some microstates. In the case of equilibrium mixed states, we explicitly build two-point quantum correlation functions, which capture the specific, superior correlations of quantum systems at finite temperature, and which are directly { accessible to experiments when correlating measurable properties}.
When non-vanishing, these correlation functions rule out a precise form of separability of the equilibrium state. In particular, we show numerically that quantum correlation functions generically exhibit a finite \emph{quantum coherence length}, dictating the characteristic distance over which degrees of freedom cannot be considered as separable. This coherence length is completely disconnected from the correlation length of the system -- as it remains finite even when the correlation length of the system diverges at finite temperature -- and it unveils the unique spatial structure of quantum correlations.     
\end{abstract}

\maketitle

\emph{Introduction.} Quantum mechanics allows for forms of correlations between separate degrees of freedom (hereafter indicated as $A$ and $B$) which are impossible in classical systems. The theoretical and experimental study of the superior forms of correlations offered by quantum mechanics represents one of the most vibrant subjects of modern quantum physics (see Refs.~\cite{Horodecki2009, Amicoetal2008, Modietal2012, Laflorencie2015, Adessoetal2016} for some recent reviews), also in sight of the technological use of such correlations as resources for quantum information processing, quantum sensing, etc. In the case of pure states, the notion of quantum correlations is unambiguously linked with entanglement, which prevents a complete description of the state of $A$ when the degrees of freedom of $B$ are traced out; pure-state entanglement can be unambiguously quantified theoretically \cite{Horodecki2009} and even measured in recent, impressive experiments \cite{Jurcevicetal2014, Fukuharaetal2015, Islametal2015}. In the case of generic mixed states, quantum correlations are much less sharply defined: a conservative definition would associate them with inseparability of the density matrix (revealing entanglement) \cite{Horodecki2009, Werner1989}; but a more general definition may be given in terms of the non-local disturbance that a local measure produces in a quantum system (captured by the so-called quantum discord \cite{Modietal2012}). Moreover, classical mixed states also possess correlations -- which can become infinitely long-ranged at critical points -- and this same form of classical correlations is found in quantum systems at finite temperature. Hence a natural question arises: is there a specific spatial structure to quantum correlations as opposed to classical ones? And can one capture this structure with the conventional tools of statistical mechanics, namely with correlation functions? 

 With this paper we answer in the positive to both questions. In particular,  in the case of equilibrium quantum systems we unveil the existence of a precise notion of quantum correlations built from elementary objects in quantum statistical mechanics, namely correlation functions and response functions. Non-zero quantum correlations among two subsystems $A$ and $B$ rule out a specific form of separability of the joint density matrix of the system, possessing a precise physical and operational meaning: the equilibrium state of the joint system cannot generically be built by simply coupling the two subsystems to some correlated, classical source of noise.     
The quantum correlation functions are explicitly calculated for two lattice boson models (hardcore bosons and quantum rotors) in two spatial dimensions, and show a striking feature: the quantum correlations decay exponentially at any finite temperature, displaying a finite \emph{quantum coherence length} $\xi_Q$, even when the (total) correlations decay algebraically as in the superfluid phase of those models. This suggests that the quantum coherence length is finite at any finite temperature in all models with short-range interactions, namely that thermal states are nearly separable over length scales far exceeding $\xi_Q$. This criterion for separability is found to be in quantitative agreement with that stemming from the skew information \cite{Chen2005}, quantum Fisher information \cite{Hyllusetal2012, Toth2012} and quantum variance \cite{FrerotR2015-2} of the most fluctuating collective observable. In contrast, the two-point quantum discord -- which can be explicitly evaluated for hardcore bosons \cite{Luo2008} -- is found to display the same decay as that of conventional correlations, seemingly adding no further information beyond the conventional diagnostics.  
 
\emph{Quantum correlation functions.}  We consider a generic quantum model with Hamiltonian ${\cal H}$, in thermal equilibrium at an inverse temperature $\beta = (k_B T)^{-1}$, and local observables $O_A$ and $O_B$ associated with two regions $A$ and $B$ of the system (taken as non-overlapping to capture correlations only). We define the two-point \emph{quantum correlation function} (QCF) between $A$ and $B$ as follows:
\begin{eqnarray}
\langle \delta O_A \delta O_B \rangle_Q  & = &  \langle \delta O_A \delta O_B \rangle - \frac{\partial\langle O_A \rangle}{\partial \lambda_B}\Big |_{\lambda_B=0}  \\
& = & \langle \delta O_A \delta O_B \rangle - \frac{1}{\beta} \int_0^{\beta} d\tau \langle \delta O_A(\tau) \delta O_B(0) \rangle~. \nonumber
\label{e.QCF}
\end{eqnarray}
Here $\langle ... \rangle = {\rm Tr}(... \rho)$ denotes the thermal average, with $\rho = e^{-\beta \cal H}/{\cal Z}$ and ${\cal Z} = {\rm Tr}(e^{-\beta \cal H})$; $\delta O = O - \langle O \rangle$ is the fluctuation with respect to the average; $\lambda_B$ is a field coupling to $O_B$ via a term $- \lambda_B O_B$ added to the Hamiltonian; and $O(\tau) = e^{\tau \cal H} O e^{-\tau \cal H}$ is the imaginary-time-evolved operator.  

 The QCF probes the difference between the conventional, two-point correlation function involving the regions $A$ and $B$, and the response function of region $A$ upon perturbing region $B$ with a field $\lambda_B$. In any classical system, these two quantities are identified via a fluctuation-dissipation relation \cite{Huangbook}, and therefore the QCF is identically zero. In a quantum system, the QCF is non-zero due to the fact that, in general, both commutators  
$[O_A,{\cal H}] $ and $[O_B,{\cal H}]$ are non-vanishing (should one of them vanish, then automatically the QCF vanishes). Hence the QCF probes how the $O_A$ and $O_B$ are \emph{jointly incompatible} with ${\cal H}$, namely how the Heisenberg's uncertainties of $O_A$ and $O_B$ on eigenstates of ${\cal H}$ (thermally populated with the Boltzmann distribution) correlate with each other. Obviously the fact that $O_A$ and $O_B$ both possess quantum uncertainty is \emph{per se} not sufficient to have a non-zero QCF: indeed if ${\cal H} = {\cal H}_A + {\cal H}_B$, so that the two subsystems $A$ and $B$ are not correlated, the QCF vanishes even if  
$[O_A,{\cal H}_A] $ and $[O_B,{\cal H}_B]$ are non-zero.  Moreover, it is easy to see that the QCF vanishes at infinite temperature $(\beta = 0)$, while it coincides with the full correlation function at $T=0$ (see Supplementary Material -- SM -- for an explicit proof \cite{SM}). 

 The above definition of QCF bears some conceptual similarity with the notion of quantum discord (QD) -- widely accepted as a most general form of quantum correlation. For an explicit definition of QD we defer the reader to the SM \cite{SM}, and to the rich literature on the subject (see Ref.~\cite{Modietal2012} for a review). Here it suffices to say that QD captures the difference between the mutual information between subsystems $A$ and $B$,
 and the 
 maximum amount of information that one can acquire on the state of $A$ by making measurements on $B$. These two informational concepts are identical in classical physics, while they differ (or are ``discordant") in the quantum context, because of the non-local disturbance that measurements on $B$ introduce on the state of $A$. 
 The QCF captures on the other hand the quantum mechanical breakdown of the identity between correlation functions and response functions, and therefore it constitutes a sort of ``fluctuation-dissipation discord" \cite{footnote}. The crucial difference between QCF and QD is the fact
 that QD is observable-independent and fully defined by the $AB$ density matrix, while the QCF is obviously observable-dependent and, most importantly, it also depends on the response to a field coupling to the observable in question. 
 
\begin{figure}[ht!]
\includegraphics[width = 0.9\linewidth]{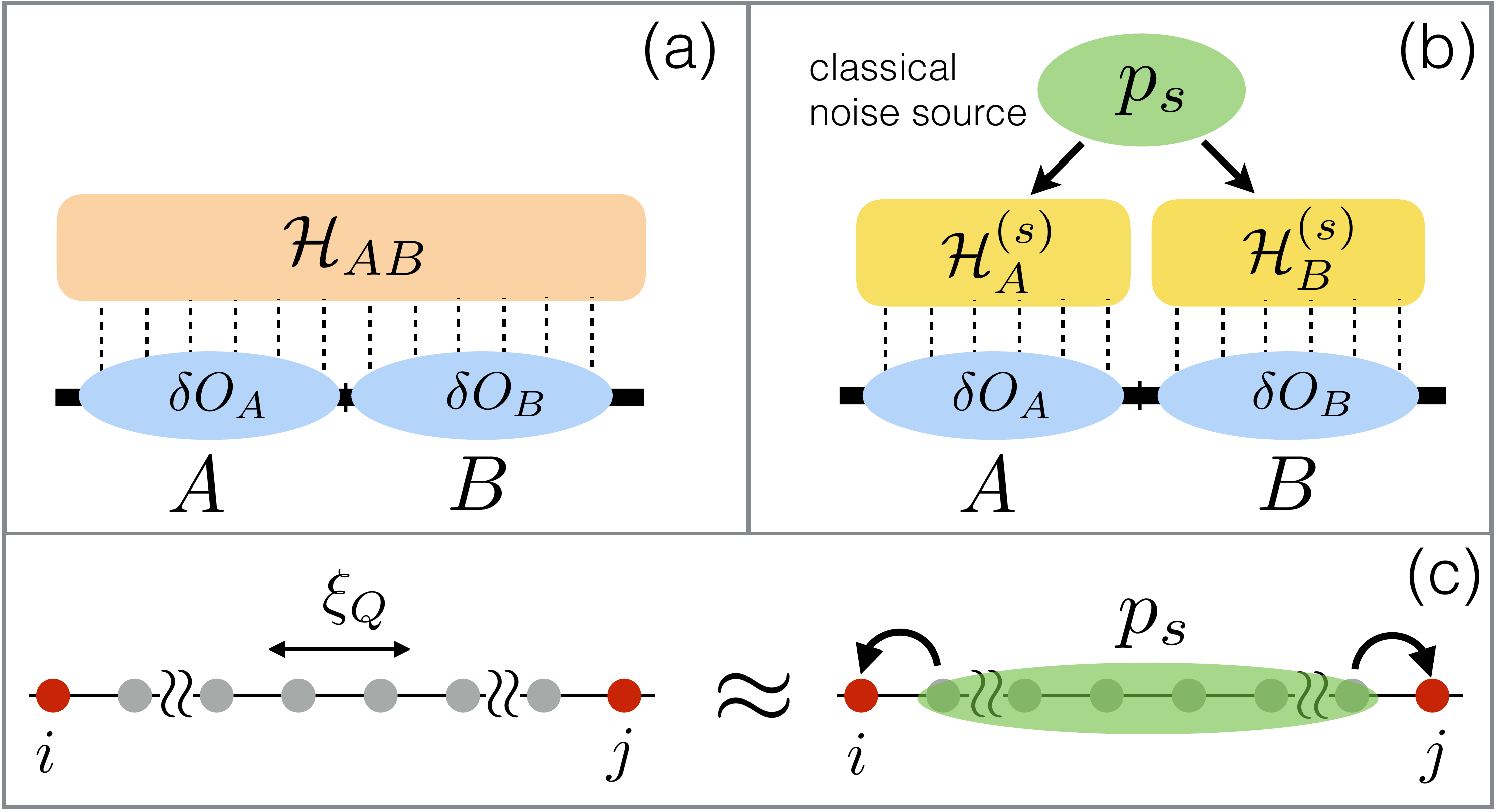} 
\caption{(a) A quantum Hamiltonian acting coherently on the subsystems $A$ and $B$ correlates quantum mechanically the uncertainties on the local observables $O_A$ and $O_B$, generating a non-vanishing quantum correlation function; (b) The opposite scenario is the one in which the two subsystems are locally in an equilibrium state governed by two separated Hamiltonians, correlated to a common classical noise source; in this case the quantum correlation function $\langle \delta O_A \delta O_B \rangle_Q$ vanishes identically; (c) When two sites $i$ and $j$ of a lattice system are separated by a distance far exceeding the quantum coherence length $\xi_Q$, the sites in between mediate their correlations in a similar way as a classical statistical field coupling the two sites.}
\label{f.separability}
\end{figure}
 
\emph{QCF and Hamiltonian-separability of equilibrium states.} A finite QCF between two subsystems implies a specific form of inseparability (namely of entanglement) between the subsystems in question. The conventional definition of separable (or classically correlated) states, due to Werner \cite{Werner1989}, amounts to assuming that the joint density matrix of the system can be written as
$\rho = \sum_s p_s \rho_A^{(s)} \otimes \rho_B^{(s)}$,
where $p_s$ is a normalized, classical distribution function weighing different factorized forms for $\rho$. In general, owing to the semi-positive definiteness of density operators, we can always write them as exponential of (effective) Hamiltonian operators:
\begin{equation}
\rho = e^{-\tilde{\cal H}} = \sum_s p_s ~e^{-{\cal H}_A^{(s)}} \otimes e^{-{\cal H}_B^{(s)}}~.
\end{equation} 
Here $\tilde{\cal H} = {\cal H} - \ln {\cal Z}$, and we have omitted any specification to the temperature, which simply defines a global scale in the spectrum of the Hamiltonian operators. 
In order to calculate the QCF, it is necessary to know explicitly how the density matrix is deformed upon applying a field term, namely upon shifting the Hamiltonian $\tilde{\cal H}$ by $-\lambda_A O_A - \lambda_B O_B$. This invites us to restrict the concept of separability to that of \emph{Hamiltonian separability}, namely to interpret ${\cal H}_{A(B)}^{(s)}$ as physical Hamiltonians which are affected linearly by the corresponding field terms. Hence we shall say that $\rho$ is Hamiltonian-separable if 
 \begin{equation}
\rho(\lambda_A,\lambda_B) = \sum_s p_s ~e^{-({\cal H}_A^{(s)}-\lambda_A O_A)} \otimes e^{-({\cal H}_B^{(s)}-\lambda_B O_B)}~.
\end{equation}
(notice that $\rho(\lambda_A,\lambda_B)$ is no longer normalized). 
Physically, Hamiltonian-separability amounts to imagining that -- as sketched in Fig.~\ref{f.separability} -- $A$ and $B$ are physical systems individually in thermal equilibrium states with Hamiltonians ${\cal H}_{A(B)}^{(s)}$, coupling both systems to a source of classical noise -- namely a classical statistical field --  whose configurations are parametrized by the parameter $s$, and which correlate $A$ and $B$ in a classical sense. A simple, practical example would be two quantum gases which are both coupled to a potential $V_s({\bm r})$, taking different configurations labeled by $s$. The fluctuations of the potential can clearly produce density-density correlations $\langle \delta n({\bm r}_A) \delta n({\bm r}_B) \rangle$ between a point in $A$ and one in $B$; but, in such a case, the density-density QCF detects the classical nature of correlations by vanishing identically.  A further example of a two-mode bosonic system is provided in the SM \cite{SM}.  

We can then state the following \underline{Theorem}: \emph{A non-zero QCF, $\langle \delta O_A \delta O_B \rangle_Q \neq 0$, for some observables $O_A$ and $O_B$ implies that $A$ and $B$ are not Hamiltonian-separable.} 
The proof of the theorem is elementary, and it can be found in the SM \cite{SM}. As a consequence, finding a finite QCF rules out that the correlations between the fluctuations of local observables $\delta O_A$ and $\delta O_B$ are generated purely by correlated classical noise.  Obviously the above result is not limited to bipartitions of the system, but it applies to $A$ and $B$ being any two subsystems in arbitrary multi-partitions of the total system.  

  \begin{figure*}[ht!]
\includegraphics[width = \linewidth]{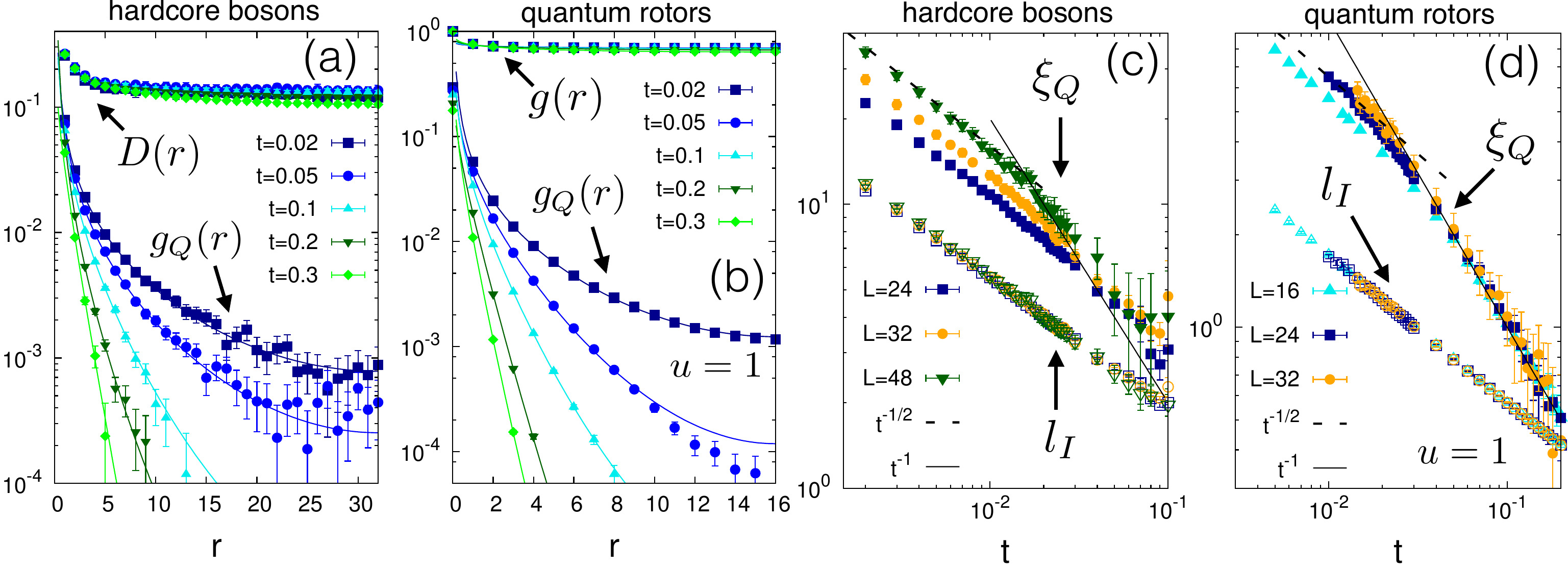}
\caption{\emph{Left panels.} Quantum correlation function $g_Q$ of Bose fields for (a) hardcore bosons with $L=64$ and different temperatures $t = k_B T/J$; (b) quantum rotors with $u = U/(2J{\bar{n}}) = 1$, $L=32$ and different temperatures $t = k_B T/ (2J{\bar n})$.
The $g_Q$ function is compared to the two-point quantum discord $D(r)$ for hardcore bosons, and to the total correlations $g(r)$ for quantum rotors. Here $r$ is the short-hand notation for $(r,0)$. Solid lines are exponential fits $A'*e^{-d(x|L)/\xi_Q} *d(x|L)^{-\eta'}$ for $g_Q$, where $d(x|L) = (L/\pi) \sin(\pi x/L)$ is the cord length. All data refer to the superfluid phase. 
\emph{Right panels.} Quantum coherence length $\xi_Q$ and inseparability length $l_I$ (see text) vs. temperature for (c) hardcore bosons; (d) quantum rotors with $u=1$. The solid and dashed lines indicate $t^{-1}$ and $t^{-1/2}$ power laws.}
\label{f.qcorr}
\end{figure*}

\emph{Quantum coherence length at finite temperature.} The QCF of any observable is readily accessible to analytical and numerical computations of all models for which we can calculate correlation and response functions. Here we exploit this property to explicitly calculate the QCF for Bose fields in two well-understood lattice boson models on a square lattice, sitting at opposite ends in the "spectrum" of possible regimes of lattice bosons: 1) hardcore bosons, described by the Hamiltonian
\begin{equation}
 {\cal H}_{\rm hc} = - J \sum_{\langle ij \rangle} \left ( \tilde{b}_i^{\dagger} \tilde{b}_j + {\rm h.c.} \right )
 \end{equation} 
 where $\langle ij \rangle$ are nearest-neighbor bonds on a lattice, and $\tilde{b}_i, \tilde{b}^{\dagger}_i$ are hardcore-boson operators anticommuting on site; and 2) quantum rotors, with Hamiltonian
 \begin{equation}
 {\cal H}_{\rm qr} = -2J\bar{n} \sum_{\langle ij \rangle} \cos(\theta_i - \theta_j) - \frac{U}{2} \sum_i \frac{\partial^2}{\partial \theta_i^2} 
 \end{equation} 
obtained as a limit of the Bose-Hubbard model with inter-particle repulsion $U$ and large, integer filling $\bar{n}$ -- in this limit, the Bose operator is decomposed into canonically conjugated amplitude and phase, $[\theta_i, n_i] = i$, and it is approximated as $b_i \approx \sqrt{n} ~e^{i\theta_i}$ in the Josephson coupling term. 

In both cases, we probe the QCF for the Bose field (or first-order QCF), namely 
\begin{equation}
g_Q(i,j) = \langle a_i^{\dagger} a_j \rangle_Q = \langle a_i^{\dagger} a_j \rangle - \frac{1}{\beta} \int d\tau \langle a_i^{\dagger}(\tau) a_j(0) \rangle
\end{equation}
where $a_i = \tilde{b}_i$ for hardcore bosons and $a_i = e^{i\theta_i}$ for quantum rotors (where we normalized the field operator by $\sqrt{\bar{n}}$). The first-order QCF can be straightforwardly calculated for hardcore bosons using quantum Monte Carlo (here in the Stochastic Series expansion formulation \cite{Sandvik1992, SyljuasenS2002}), and for quantum rotors using path-integral Monte Carlo \cite{Wallinetal1994}. First-order correlations are the dominant ones in the above models, transitioning from exponentially decaying (in the normal phase) to algebraically decaying (in the superfluid phase) at a Kosterlitz-Thouless (KT) transition occurring at temperature $T_{\rm KT}$ -- associated with the divergence of the correlation length $\xi(T)$. Given the special role of first-order correlations, one may naturally expect that the first-order QCF are also the dominant ones among all QCFs -- something which we verified explicitly in our numerical simulations. Hence, according to the previous section, the first-order QCF captures the degree of (Hamiltonian-)separability between two sites. 

Fig.~\eqref{f.qcorr} shows the first-order QCF in the superfluid phase of hardcore bosons and quantum rotors. It is striking to observe that, in both cases, the QCF lays orders of magnitudes below the total correlation function $g(i,j) = \langle a_i^{\dagger} a_j \rangle$ down to very low temperatures. Most importantly, it decays exponentially at \emph{all} finite temperatures, revealing the existence of a characteristic quantum coherence length $\xi_Q$ which is completely insensitive to { the divergent correlation length associated with the superfluid phase}. We extract systematically this coherence length on $L\times L$ lattices from the (Lorentzian) width of the $k=0$ peak in the ``quantum" momentum distribution, $n_{\rm Q}({\bm k}) = \frac{1}{L^2} \sum_{ij} e^{i {\bm k}\cdot ({\bm r}_i -{\bm r}_j)}~ g_Q(i,j)$, which is assumed to behave as $n_Q(2\pi/L,0) \approx n_Q(0)/[1+(2\pi\xi_Q/L)^2]$. 
The temperature dependence of the $\xi_Q$ so extracted is shown in Fig.~\ref{f.qcorr}(a)-(b), where we observe that $\xi_Q$ diverges upon lowering the temperature. The asymptotic temperature dependence of $\xi_Q$, while presumed to follow a power law ($\xi_Q \sim T^{-\alpha}$) is difficult to extract from the numerics  -- where one can clearly observe crossovers between at least two temperature behaviors -- but it can be predicted analytically (\emph{e.g.} on the basis of spin-wave theory), and it shall be discussed in a forthcoming publication \cite{Rancon_inpreparation}. 

The quantum coherence length sets the characteristic scale beyond which two subsystems can be considered as nearly Hamiltonian-separable -- in explicit physical terms, when $g_Q(i,j) \ll 1$, the correlations between the two points $i$ and $j$ could have been prepared by coupling two independent subsystems (containing sites $i$ and $j$ respectively) to the same source of classical noise. Obviously this source does not exist physically, but one can consider the degrees of freedom spatially separating the sites $i$ and $j$ as the effective ``classical bus" for correlations among the two sites -- classical because the distance between $i$ and $j$ exceeds the quantum coherence length (see Fig.~\ref{f.separability}(c)). A completely alternative probe of separability (in the general sense of Ref.~\cite{Werner1989}) has been recently introduced for lattice systems based on the quantum Fisher information \cite{Hyllusetal2012, Toth2012}, skew information \cite{Chen2005} and quantum variance \cite{FrerotR2015-2} of a collective observable. In particular the last criterion (detailed in the SM \cite{SM}) defines a minimal ``inseparability length" $l_I$ (corresponding to the \emph{minimal} linear size of clusters into which the density matrix can be separated, based on the quantum fluctuations of a collective observable), which takes the form $l_I = \sqrt{2n_Q(k=0)}$ for hardcore bosons and  $l_I = \sqrt{n_Q(k=0)/2}$ for quantum rotors.  Fig.~\ref{f.qcorr}(c),(d) shows that $l_I < \xi_Q$ (as expected from the definition of $l_I$), and that the two lengths display a very similar temperature dependence at low temperatures. These findings strengthen the interpretation of $\xi_Q$ as the characteristic length beyond which two subsystems can be considered as nearly (Hamiltonian-)separable. 

\emph{Quantum correlation functions vs. quantum discord.} The central claim of our paper is that the QCF captures an essential form of quantum correlation between local observables belonging to distinct subsystems of an extended quantum system in thermal equilibrium. This immediately calls for a comparison with quantum discord (QD), which is a general, observable-independent definition of quantum correlations. The comparison is immediate in the case of hardcore bosons, mapping onto $S=1/2$ spins, for which a calculable expression for two-point QD exists \cite{Luo2008, Sarandy2009}. In the limit $|{\bm r}_i - {\bm r}_j|\to \infty$, and in the absence of spontaneous symmetry breaking, the asymptotic two-point QD, $D(i,j)$, can be easily shown to take the form (see \cite{SM})
\begin{equation}
D(i,j) \approx (C_1^2 + C_2^2)/(2\log 2)~
\end{equation}
where  $C_1 =  g(i,j)/2$ and  $C_2  = \langle n_i n_j \rangle - \langle n_i \rangle \langle n_j \rangle$ are the correlation functions of the system, decaying to zero at large distance. Therefore for 2$d$ hardcore bosons the two-point QD decays algebraically throughout the superfluid phase and exponentially only in the normal phase, in a similar way as ordinary correlations do -- and it is singular at the KT transition, even though this transition is uniquely driven by thermal fluctuations. This is to be contrasted with the QCF, not bearing any signature of the KT transition \cite{SM}. { The dramatic difference between QD and QCF, and the sensitivity of QD to classical critical phenomena, suggests that the notion of quantum correlations attributed to QD should be critically re-examined.} The sensitivity of the two-point QD to ordinary correlations can be simply traced back to its definition in terms of the reduced two-point density matrix $\rho_{ij}$ which is in turn fully expressed through correlation functions. On the other hand, the QCF depends on the reduced density matrix \emph{and} its deformation upon applying a field at site $i$ (or $j$) -- in quantum statistical mechanics one would say that it depends on the full structure of imaginary-time propagators \cite{NegeleO1998}, which reduce to correlation functions at equal times. { As a consequence the QCF provides information beyond that contained in ordinary correlations and in the two-point QD.}

\emph{Conclusions.}  We have introduced the concept of quantum correlation functions (QCFs) for equilibrium quantum many-body systems, capturing the part of correlations among two subsystems that cannot be ascribed to the coupling with a common classical source of noise. QCFs unveil the existence of a finite quantum coherence length, completely independent of the correlation length in the system, beyond which quantum correlations are exponentially suppressed. QCFs are uniquely defined in terms of measurable quantities (full correlation and response functions) and therefore they are directly accessible to experiments, as well as to analytical and numerical calculations on large-scale systems. While we investigated them in the context of bosonic models, QCFs are immediately generalizable to fermionic systems. Their systematic investigation opens the path towards a deep understanding of quantum correlations in realistic thermal states, especially interesting when considering \emph{e.g.} the quantum critical ``fan" at $T>0$ for many-body systems displaying a zero-temperature quantum phase transition \cite{Sachdevbook}.   
   
\emph{Acknowledgements.} We thank I. Fr\'erot and A. Ran\c con for invaluable discussions and suggestions, as well as a careful reading of the manuscript. This work is supported by ANR (``ArtiQ" project).

\vspace{2cm}

\begin{center}
{\bf
SUPPLEMENTARY MATERIAL\\
``Quantum correlations, separability and quantum coherence length in equilibrium many-body systems"} 
\end{center}

\begin{appendix}   

\section{Quantum correlation functions and total correlation functions coincide at $T=0$}

To prove that $\langle \delta O_A \delta O_B \rangle_Q = \langle \delta O_A \delta O_B \rangle$ coincide in the ground state of a given system, we prove that the imaginary-time correlation function $\langle \delta O_A (\tau) \delta O_B(0) \rangle_0$, calculated for the ground state of the system, is a function whose absolute value decreases to zero for $\tau \to \infty$. This in turn implies necessarily that the integral 
\begin{equation}
\left | \int_0^{\beta} d\tau  \langle \delta O_A (\tau) \delta O_B(0) \rangle_0 \right | 
\leq    \int_0^{\beta} d\tau  \left | \langle \delta O_A (\tau) \delta O_B(0) \rangle_0 \right |
\end{equation}
can only grow slowlier than linearly with $\beta$, so that it vanishes when normalized by $\beta^{-1}$, as in Eq.~(1) of the main text. 

Making use of the basis $|n\rangle$ of eigenstates of ${\cal H}$ with eigenenergies $E_n$, and assuming that the system admits a non-degenerate ground state $|0\rangle$, one has that
\begin{eqnarray}
\left | \langle \delta O_A (\tau) \delta O_B(0) \rangle_0 \right |  = \left | \sum_{n >0} \langle 0 | \delta O_A | n \rangle \langle n | \delta O_B | 0 \rangle e^{-\Delta E_n \tau} \right | \nonumber \\
\leq \sum_{n >0} \left | \langle 0 | \delta O_A | n \rangle \langle n | \delta O_B | 0 \rangle \right | e^{-\Delta E_n \tau} ~~~~~~
\end{eqnarray}
where $\Delta E_n = E_n - E_0 > 0$, and the ground-state term disappears because $\langle 0 | \delta O_{A(B)} | 0 \rangle  = 0$ by construction. Hence the imaginary-time correlation function is a sum of exponentially decreasing terms, and it decreases to zero in the limit $\tau \to \infty$. The case of a degenerate ground state is somewhat pathological, { because in that case the equilibrium state at $T=0$ is not well defined}, but one can easily circumvent this difficulty physically, by lifting the degeneracy with an infinitesimal perturbation, and remark that the identity between quantum and total correlation function at $T=0$ is completely independent of the perturbation.  
 
\vspace{1cm}

\section{Hamiltonian-separability theorem}

In this section we prove the theorem announced in the main text. Considering a Hamiltonian-separable density matrix for subsystems $A$ and $B$, $\rho(\lambda_A,\lambda_B)$ as in Eq.~(3) of the main text, and the corresponding partition function ${\cal Z}(\lambda_A,\lambda_B) = {\rm Tr} \rho(\lambda_A,\lambda_B)$, we have that 
\begin{widetext}
\begin{eqnarray}
\frac{\partial \langle O_A \rangle}{\partial \lambda_B} \Big|_{\lambda_{A(B)}=0}  & = & 
\frac{1}{{\cal Z}(0,0)} \sum_s p_s ~{\rm Tr}_A\{O_A e^{-{\cal H}_A^{(s)}}\} \frac{\partial}{\partial \lambda _B} {\rm Tr}_B \{ e^{-({\cal H}_B^{(s)}-\lambda_B O_B)} \}  ~\Big|_{\lambda_{B}=0} - \langle O_A \rangle \frac{1}{{\cal Z}(0,0)} \frac{\partial {\cal Z} (0,\lambda_B)}{\partial \lambda_B}~\Big|_{\lambda_{B}=0}  \nonumber \\
&=& \langle O_A O_B \rangle - \langle O_A \rangle \langle O_B \rangle~.
\end{eqnarray}
\end{widetext}
As a consequence, Hamiltonian separability implies the vanishing of all QCFs $\langle \delta O_A \delta O_B \rangle_Q$, and the presence of at least one non-vanishing QCF negates Hamiltonian separability.  

\section{Example of Hamiltonian (in)separability: two-mode boson system}

To capture the essence of Hamiltonian separability, one can consider the situation of a two-mode bosonic system, with Hamiltonian
\begin{equation}
{\cal H} = -J (b_A^\dagger b_B + {\rm h.c.}) + U (b_A^\dagger)^2 b_A^2  + U (b_B^\dagger)^2 b_B^2
\end{equation}
containing a hopping term ($J$) and a repulsion ($U$) term. In this system the field correlations $\langle b_A^\dagger b_B \rangle$ have a non-zero quantum component, established by the coherent hopping term, and the density matrix is not Hamiltonian-separable. 

The Hamiltonian-separable version of this system would have a Hamiltonian ${\cal H} = {\cal H}_A + {\cal H}_B$, where  
\begin{eqnarray}
{\cal H}_A(\Psi, \Psi^*)  & = & -(\Psi^* b_A + {\rm h.c.})  + U (b_A^\dagger)^2 b_A^2 \nonumber \\
 {\cal H}_B(\Psi, \Psi^*)  & = & -(\Psi^* b_B + {h.c.}) + U (b_B^\dagger)^2 b_B^2
\end{eqnarray} 
dependent on a complex classical field $\Psi$, and a corresponding density operator 

\begin{equation}
\rho = \frac{1}{\cal Z} \int \frac{d\Psi d\Psi^*}{2\pi i} P(\Psi, \Psi^*)  ~e^{-{\cal H}_A} \otimes e^{-{\cal H}_B}~.
\label{e.rhopsi}
\end{equation} 
 
 In this case the field correlations $\langle b_A^\dagger b_B \rangle$ are induced uniquely by the classical field term, and do not admit a quantum part, namely $\langle b_A^{\dagger} b_B \rangle_Q = 0$.   
 {
 Indeed, introducing the notation
 \begin{equation}
 \langle b_{A(B)} \rangle_{\Psi}  =  {\rm Tr} \left[ b_{A(B)} e^{-{\cal H}_A(\Psi, \Psi^*)}\right ]  
\end{equation}
one has that
\begin{equation}
\langle   b_A^\dagger b_B \rangle = \frac{1}{\cal Z} \int \frac{d\Psi d\Psi^*}{2\pi i} P(\Psi, \Psi^*)    \langle b^{\dagger}_{A} \rangle_{\Psi} \langle b_{B} \rangle_{\Psi}~\neq 0.
\end{equation}
Even if none of the factorized density matrices which are superposed in Eq.~\label{e.rhopsi} possess off-diagonal terms coupling $A$ and $B$, and averages $\langle b_A^\dagger b_B \rangle_{\Psi}$ factorize, correlations do exist between the average values $\langle b_{A(B)} \rangle_{\Psi}$  via the common coupling to the $\Psi$ field. On the other hand, quantum correlations vanish as a consequence of the theorem discussed in the previous section.}

\section{Quantum variance criterion for inseparability}   

\subsection{Quantum variance on separable states}
  
   A criterion for inseparability, which can be tested quantitatively on generic many-body systems, is based on the \emph{quantum variance} (QV) of a collective operator \cite{FrerotR2015-2}. Given a generic operator ${O}$ its quantum variance on a  generic state $\rho$ is defined as 
  \begin{equation}
  \langle \delta^2 O \rangle_Q = \langle O^2 \rangle - \frac{1}{\beta}\int_0^{\beta} d\tau ~\langle O(\tau) O(0) \rangle~.
  \label{e.QV}
  \end{equation} 
  The definition of the QV is natural for thermal states, and completely accessible whenever the model of interest is solvable (analytically or numerically) at equilibrium.  But it remains valid also for generic non-thermal states, given that a generic density operator can always be written as $\rho= \exp(-\beta{\cal H})/{\rm Tr}[\exp(-\beta{\cal H})]$ for some (effective) Hamiltonian ${\cal H}$ and inverse temperature $\beta$: this in turn allows generically to define the imaginary-time evolved operator $O(\tau)$. 
  
  In the following we shall focus on a collective operator ${O} = \sum_i o_i$, which is the sum of local operators $o_i$ with a bounded spectrum in $[o_{\rm min}, o_{\rm max}]$. 
 And we consider a state $\rho$ which can be separated into clusters of maximal size $p$ (or $p$-separable), namely which admits the separable form
  \begin{equation}
  \rho = \sum_s p_s \otimes_c \rho_c^{(s)}
  \label{e.pprod}
  \end{equation}
  where $\rho_c^{(s)}$ is the density matrix for a single cluster. 
The QV of a collective operator satisfies a fundamental bound for separable states of the form Eq.~\eqref{e.pprod} thanks to two of its main properties, namely: 1) the QV is convex; 2) the QV is upper bounded by the total variance, $\langle \delta^2 O \rangle_Q \leq \langle \delta^2 O \rangle = \langle O^2 \rangle - \langle O \rangle^2$. Hence for separable states it admits the bound 
\begin{eqnarray}
\langle \delta^2 O \rangle_Q & \leq & \sum_s ~p_s ~ \sum_c  \langle \delta^2 O_c \rangle_{Q,s} \nonumber \\
&\leq &  \sum_s ~p_s ~ \sum_c  \langle \delta^2 O_c \rangle_s 
\end{eqnarray}
where ($\langle \delta^2 O_c \rangle_{Q,s}$) $\langle \delta^2 O_c \rangle_{s}$ is the (quantum) variance of the cluster operator $O_c = \sum_{i \in c} o_i$ on the cluster state $\rho_c^{(s)}$. Here we used the above cited properties of the QV, as well as the absence of any correlation between clusters within the factorized state $\otimes_c \rho^{(s)}_c$.  Considering for simplicity a partition of the system of total size $N$ into identical clusters of size $p$ (such that $N/p$ is an integer), the variance of the observable $O_c$ is easily upper-bounded as
\begin{equation}
\langle \delta^2 O_c \rangle \leq \frac{p^2}{4} (o_{\rm max} - o_{\rm min})^2  
\end{equation} 
where the bound corresponds to a bimodal distribution for the observable $O_c$ with values $p o_{\rm max}$ and $p o_{\rm min}$ both having probability 1/2.  
As a consequence, for all $p$-separable states, the QV satisfies the bound: 
\begin{equation}
\langle \delta^2 O \rangle_Q \leq \frac{N p}{4} ~(o_{\rm max} - o_{\rm min})^2~.  
\label{e.pbound}
\end{equation}
The above bound is obviously a necessary but not sufficient condition for $p$-separability, namely states which are not $p$-separable but are $p'$-separable with $p'>p$ (up to $p' = \infty$) may still comply with the bound.  

The same bound applies to other quantities discussed previously in the literature, namely the Wigner-Yanase skew information \cite{WignerY1963, Chen2005} and the quantum Fisher information \cite{BraunsteinC1994,Toth2012, Hyllusetal2012, PezzeS2014, Haukeetal2016} -- from which the present discussion is strongly inspired. As a matter of fact, the QV represents a tight lower bound to both the skew information and the quantum Fisher information \cite{FrerotR2015-2}. While the skew and quantum Fisher information would tighten the bound in Eq.~\eqref{e.pbound} for $p$-separable states, they are unfortunately not accessible to large-scale quantum many-body calculations.  

\subsection{``Quantum momentum distribution" and separability}

{
The bound in Eq.~\eqref{e.pbound} allows to use the quantum variance as a witness of entanglement. Indeed, given a thermal state with quantum variance $\langle \delta^2 O \rangle_Q$, in order to approximate it with a $p$-separable state one needs to use clusters with $p$ at least taking the value which saturates the bound of Eq.~\eqref{e.pbound}, namely $p \geq p_{\rm min} = 4\langle \delta^2 O \rangle_Q/N $. Hence the quantum variance witnesses entanglement among at least $p_{\rm min}$ sites.} Considering then the local observables $o_i$ with unit spectral width, $\Delta o = o_{\rm max} - o_{\rm min} = 1$, which maximize the quantum variance of the corresponding collective observable $O$ among all collective observables, then for a $d$-dimensional system one can define an \emph{inseparability length} $l_I$ as 
\begin{equation}
l_I = \left[ \frac{4}{N} \sup_{O: \Delta o = 1} ~\langle \delta^2 O \rangle_Q \right]^{1/d}~.
\label{e.lI}
\end{equation}
This length indicates the \emph{minimal} linear size of clusters building a separable state of the kind of Eq.~\eqref{e.pprod}, which is compatible with the maximum quantum variance of collective observables. It is therefore to be considered as a lower bound to the length beyond which two subsystems can be considered as effectively separable in the state of the system. Hence it is meaningful to compare it to the quantum coherence length $\xi_Q$, which is the natural (Hamiltonian-)separability length, and to which $l_I$ may be expected to act as a lower bound. 

The length $l_I$ is easily identified in the case of the two bosonic models of interest in this work. Indeed one can maximise the quantum variance of a collective observable with $\Delta o = 1$ by considering:
 1) for hardcore bosons, $o_i = (\tilde{b}_i + \tilde{b}_i^{\dagger})/2$; 2) for quantum rotors, $o_i = [\cos(\theta_i) + \sin(\theta_i)/(2\sqrt{2})$. { In both cases the quantum correlation function $\langle o_i o_j \rangle_Q$ is proportional to the quantum field correlation function, namely $\langle o_i o_j \rangle_Q = g_Q(i,j)/2$ (for hardcore bosons at half filling) and $\langle o_i o_j \rangle_Q = g_Q(i,j)/8$ (quantum rotors).} As a consequence, 
 the corresponding quantum variance is related to the $k=0$ peak in the ``quantum momentum distribution" $n_Q(k)$, namely $\langle \delta^2 O \rangle_Q  = N n_Q(0)/2$ (hardcore bosons at half filling) and $\langle \delta^2 O \rangle_Q  = N n_Q(0)/8$
 (quantum rotors).  { Given that the $g_Q(i,j)$ is the dominant quantum correlation function, and it is positive definite, its integral will give the dominant quantum variance among all observables, as requested in Eq.~\eqref{e.lI}.} The resulting inseparability length $l_I \sim \sqrt{n_Q(k=0)}$ is then defined in the main text. 
 
 On general grounds, one can establish a scaling relationship between the two lengths $l_I$ and $\xi_Q$ based on the fact that they are derived from the same function $n_Q(k)$, or, alternatively, its inverse Fourier transform, $g_Q(i,j)$. 
 Indeed, one can expect the quantum correlation to decay as:
  \begin{equation}
  g_Q(r) \sim \frac{e^{-r/\xi_Q}}{r^{d-2+\tilde\eta}}
  \end{equation}
  (which is verified by the fits of the numerical data in Fig.~2 of the main text). Therefore, integrating $g_Q(r)$ one obtains  
  \begin{equation}
  n_Q(k=0) \sim \int_a^{\infty} dr ~r^{d-1} g_Q(r)  \sim \xi_Q^{2-\tilde\eta} \int_{a/\xi_Q}^{\infty} dx \frac{e^{-x}}{x^{\tilde\eta-1}}
  \end{equation}
  where $a$ is the lattice spacing. Under the assumption that $\xi_Q/a \gg 1$, the integral loses its dependence on $\xi_Q$, and hence one obtains the scaling relation 
  \begin{equation}
  l_I \sim [n_Q(\bm k=0)]^{1/d} \sim \xi_Q^{\frac{2-\tilde \eta}{d}}.
  \end{equation}
  Hence the temperature dependence of the two lengths is generically different unless $\phi = (2-\tilde{\eta})/d = 1$. The data in Figs.~2(c-d) of the main text suggest that, for the models of interest, $1/2 \lesssim \phi \lesssim 1$ in the low-temperature regime of $\xi_Q/a \ll 1$.  In general one can expect that $\phi \leq 1$, so that the inequality $l_I \leq \xi_Q$ holds for $T\to 0$, where both quantities diverge.

\section{Quantum discord and correlation functions}
 
 The discussion provided in the main text essentially identifies the concept of equilibrium quantum correlations with that of (Hamiltonian) separability -- for the fundamental reason that, in the context of equilibrium Bose fields, Hamiltonian separability implies the vanishing of quantum correlation functions, and viceversa. On the other hand, the concept of quantum correlations has been vastly extended with respect to that of entanglement and separability with the introduction of \emph{quantum discord} \cite{OllivierZ2001,HendersonV2001,Modietal2012}, which fundamentally expresses to what extent a density matrix violates an identity valid for classical, joint probability distributions of several variables. This violation stems from the non-local disturbance that local measurements create in quantum mechanics, and which is present even in the case of separable states. 
 
 \subsection{Definition of quantum discord}

Even though the quantum discord is thoroughly discussed in the existing literature, we find it useful to provide a short description of its definition and physical meaning in the following.  In the spirit of two-point quantum correlations, which this work focuses on, we shall isolate two sites $i$ and $j$ in the lattice, and define the reduced density matrices $\rho_i$, $\rho_j$ and $\rho_{ij}$ for the single sites $i$ and $j$, and for the two-site compound $ij$, respectively. The total amount of correlations (be them of classical or quantum origin) among the two sites is generally expressed via the mutual information
\begin{equation}
I(i,j) = S(\rho_i) + S(\rho_j) - S(\rho_{ij})
\end{equation}    
where $S(\rho) = - {\rm Tr} (\rho \log_2 \rho)$ is the von Neumann entropy. The mutual information expresses the ``missing" entropy in the compound state due to correlations in the fluctuations: namely, there exists information on $i$ which can be gained by making observations on $j$, and viceversa. Indeed $S(\rho_{ij} | \rho_j) = S(\rho_{ij}) -  S(\rho_j)$ is the entropy of $ij$ conditioned on the knowledge of the state of $j$, and the fact that this entropy is less than that of $\rho_i$ (ignoring completely $j$) implies the existence of correlations between $i$ and $j$, which provide information on $i$ when measuring $j$. 

This observation invites to analyze the density matrix conditioned on measurements on site $j$. Considering the local observable $O_j$  on site $j$ with eigenvalues $o^{(j)}_k$ and projectors ${\cal P}_k^{(j)}$ on the associated eigenspaces, one can define
\begin{equation}
\rho_{ij,k} = \frac{1}{p_k} \left(\mathbb{I}_i \otimes {\cal P}_k^{(j)}\right ) ~\rho_{ij} ~\left(\mathbb{I}_{i} \otimes {\cal P}_k^{(j)} \right) 
\end{equation}
where $p_k = {\rm Tr}\left[\left(\mathbb{I}_{i} \otimes {\cal P}_k^{(j)}\right ) \rho \left(\mathbb{I}_{i} \otimes {\cal P}_k^{(j)} \right)  \right]$. $\rho_{ij,k}$ is therefore the compound density matrix of sites $ij$ conditioned upon the outcome $o_k^{(j)}$ of the measurement of the observable $O_j$. The compound entropy conditioned on the measurement of the observable $O_j$ can be therefore expressed as
\begin{equation}
S(\rho_{ij}|O_j) = \sum_k p_k S(\rho_{ij,k})~,
\label{e.qcondition}
\end{equation}
which expresses the average entropy that the system has after a measurement of the observable $O_j$ -- averaged over all the possible outcomes of the measurement, with their \emph{a priori} probabilities $p_k$. In a classical system the $p_k$ would be the statistical weights of the configurations of site $j$, and therefore Eq.~\eqref{e.qcondition} would represent the entropy of $ij$ conditioned upon the knowledge of $j$, $S(\rho_{ij} | \rho_j)$. In a quantum-mechanical system, this is no longer the case, because measurements on $j$ not only give information on $i$, but also perturb its state. The amount by which measurements on $j$ perturb $i$ is then quantified by the quantum discord
\begin{equation}
D(i,j) = I(i,j) - C(i,j)
\end{equation}
where 
\begin{equation}
C(i,j) =  S(\rho_i) - {\rm max}_{O_j} S(\rho_{ij}|O_j) 
\label{e.class}
\end{equation}
expresses the (so-called) ``classical" correlations, namely the maximum amount of information that can be gained on $i$ by making measurements on $j$. The function $D(i,j)$ captures the fundamental discrepancy (or ``discord") between the entropy associated with the correlations among sites $i$ and $j$, and the maximal information that one can gain on $i$ by making projective measurements on $j$: the latter does not saturate the former because local measurements disturb the state and they reduce correlations between $i$ and $j$.  

 In summary, seen as a generalized correlation function, $D(i,j)$ probes how much a measurement on $i$ can affect the state of $j$. Even for states in which two subsystems are separable, the measurement on one system can affect the state of the other (this is true when the factorized density matrices $\rho^{(s)}_{A(B)}$ in the separable form do not commute with each other). Hence quantum discord can be non-zero even in the presence of separability. Therefore one can naturally expect the range of $D(i,j)$ to extend much further than that of the quantum correlation functions. Quantifying discord for generic degrees of freedom is in general a hard problem, due to the maximization operation implied in Eq.~\eqref{e.class}. Nonetheless two-point quantum discord admits a computable form in the case of $S=1/2$ spins \cite{Luo2008, Sarandy2009}, to which hardcore bosons reduce under spin-boson mapping: ${\tilde b}_i = S_i^-$, ${\tilde b}_i^{\dagger} = S_i^+$, and $n_i -1/2 = S_i^z$. Hence this enables the quantitative investigation of quantum discord for a lattice bosons problem, as already exploited in the recent literature \cite{Sarandy2009}.

\begin{figure}[ht!]
\includegraphics[width = \linewidth]{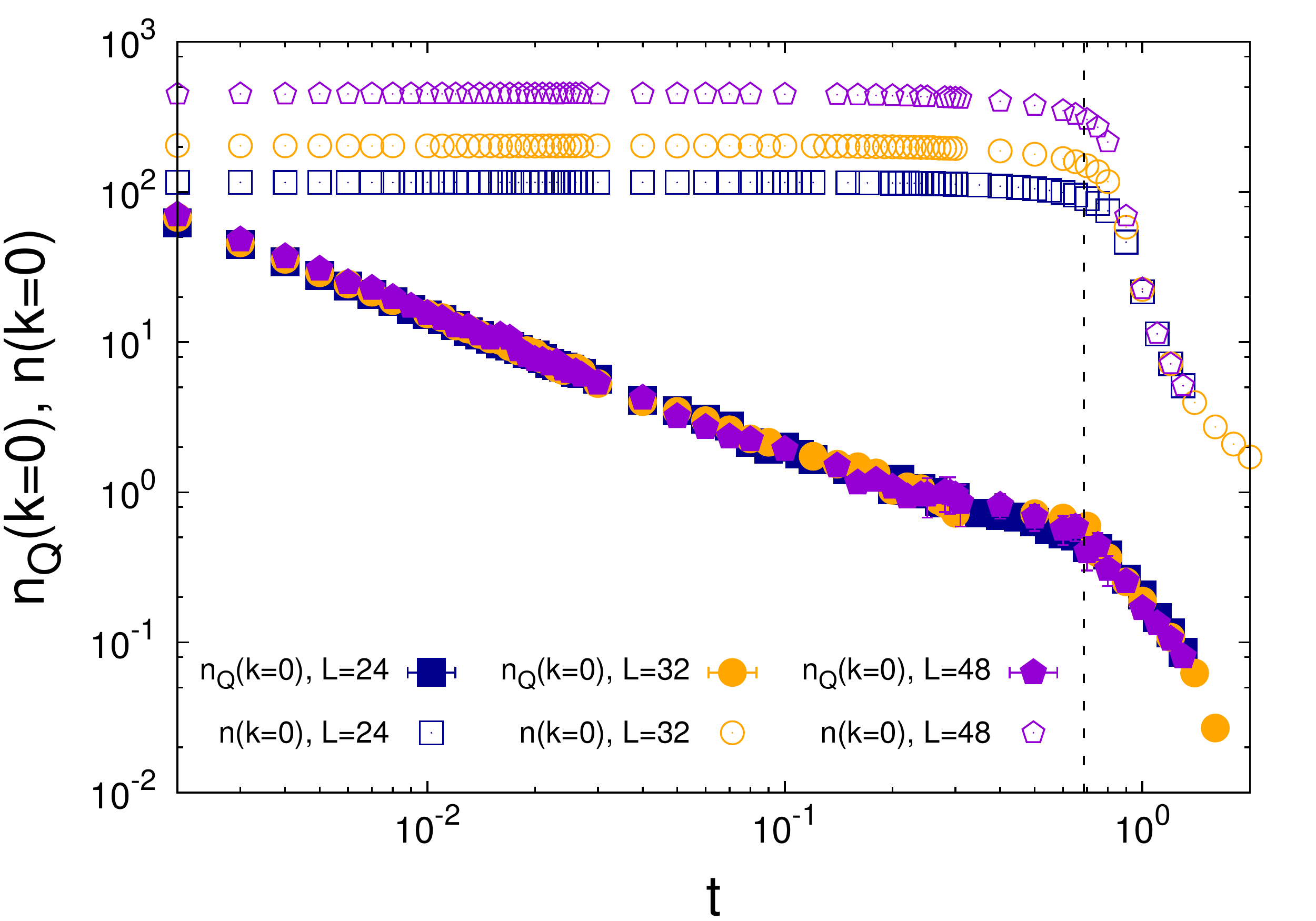} 
\caption{Zero-momentum peak in the total and ``quantum" momentum distribution for 2$d$ hardcore bosons. Full symbols refer to $n_Q(k=0)$, while open ones to $n(k=0)$. { The latter quantity diverges with system size as $L^{2-\eta(t)}$ below the 
 reduced temperature $t =  k_B T/J  = t_{\rm KT} \approx 0.685$ \cite{KawashimaH1998} (marked by a dashed line), corresponding to the Kosterlitz-Thouless critical temperature. Here $\eta(t)$ goes continuously from 0 at $t=0$ to $1/4$ at $t_{\rm KT}$. On the other hand, any size dependence  in $n_Q(k=0)$ is only appreciable at very low temperature.}}
\label{f.enk}
\end{figure}

\subsection{Two-point quantum discord for hardcore bosons}

As discussed in Ref.~\cite{Luo2008}, two-qubit quantum discord can be completely expressed in terms of two-point correlation functions. In the presence of $U(1)$ symmetry, only two correlation functions are relevant: 
\begin{eqnarray}
C_1 & = & \langle S_i^x S_j^x \rangle =  \langle S_i^y S_j^y \rangle = \frac{1}{2} \langle b_i^{\dagger} b_j \rangle \nonumber \\
C_2 & = & \langle S_i^z S_j^z \rangle =  ( \langle n_i n_j \rangle - \langle n_i \rangle \langle n_j \rangle)~.
\end{eqnarray}
 When the system further possesses a $\mathbb{Z}_2$ symmetry along the $z$ axis, the mutual information takes the form
 \begin{equation}
 I(i,j) = 2 + \sum_{n=0}^3 \lambda_n \log_2 \lambda_n
 \end{equation} 
 where $\lambda_0 = 1/4- 2C_1 - C_2$, $\lambda_1 = \lambda_2 = 1/4 - C_2$, $\lambda_3 = 1/4 + 2C_1 - C_2$. On the other hand, the ``classical" correlations admit a closed form as 
 \begin{equation}
 C(i,j) = \frac{1-C}{2} \log_2(1-C) + \frac{1+C}{2} \log_2(1+C)
 \end{equation}
 where $C = 4 \max (C_1,C_2)$.

Considering extended quantum systems with decaying correlations, it is very instructive to investigate the decay of the two terms entering in the quantum discord. In the limit $C_1, C_2 \ll 1$, valid for $|{\bm r}_i - {\bm r}_j|\to \infty$ at $T>0$, expanding the logarithms up to second order one can straightforwardly show that 
\begin{eqnarray}
I(i,j) & \approx & \frac{1}{2\log 2} \left ( 2C_1^2 + C_2^2) + {\cal O}(C_1^3, C_2^3 \right ) \nonumber \\
C(i,j) & \approx & \frac{C^2}{2\log 2} + {\cal O}(C^3) 
\label{e.asymptotics}
\end{eqnarray}
In the case of the equilibrium state of hardcore bosons $C_1 > C_2$ at all temperatures (namely first-order correlations dominate over second-order ones), and therefore the quantum discord becomes simply 
\begin{equation}
D(i,j) \approx \frac{1}{2\log 2} (C_1^2 + C_2^2)
\end{equation}
as announced in the main text. Therefore the quantum discord, being asymptotically proportional to the square of the correlation functions, has their exact same range. At variance with the quantum correlation function, it does not provide any further information about the system that (conventional) correlation functions do not already possess. Despite its quantum nature, two-point quantum discord experiences all the singular features of correlation functions at \emph{thermal} transitions: in the case at hand, it has a slow algebraic decay (as $|{\bm r}_i - {\bm r}_j|^{-2\eta}$) below the KT critical temperature (as shown in Fig.2(a) of the main text), while it abruptly changes to exponentially decaying behavior at the KT temperature. On the other hand, quantum fluctuations and correlations are not supposed to exhibit singularities at thermal phase transitions. Hence the behavior of the two-point quantum discord is in sharp contrast with that of the quantum correlation function (decaying exponentially at all finite temperatures). Fig.~\ref{f.enk} shows indeed that the ``quantum" momentum distribution $n_Q(k=0)$, namely the integral of the quantum correlation function, does not exhibit any singularity at the KT transition of hardcore bosons (occurring for $t = k_B T/J \approx 0.685$ \cite{KawashimaH1998}). On the other hand, at the KT critical temperature the integral of the total correlations, $n(k=0)$, exhibits a well-known divergence -- and the same faith is shared with the integral of the 2-point quantum discord. 

 Finally, Eq.~\eqref{e.asymptotics} shows that the identification of $C(i,j)$ with classical correlations is somewhat problematic -- when one endows ``correlations" with the meaning generally attributed to it in statistical mechanics. Indeed $C(i,j)$ is obviously non-zero at $T=0$, where on the other hand a classical system has no correlations at all. Hence the ``classical" attribute to $C(i,j)$ has to be taken in the informational sense (that $C(i,j)$ would classically be equal to the mutual information), and not in the physical sense (namely that $C(i,j)$ has the same properties as a correlation function of a classical system).

\end{appendix}

\bibliography{AFMC}

\end{document}